\documentclass{aa}  

\usepackage{graphicx}
\usepackage{txfonts}
\begin{document}

   \title{Occurrence and persistence of magnetic elements in the quiet Sun}


   \author{F. Giannattasio
          \inst{1,*}
          \and
          F. Berrilli\inst{2}
          \and
          G. Consolini\inst{1}
          \and
          D. Del Moro\inst{2}
          \and
          M. Go{\v{s}}i{\'c}\inst{3}
          \and
          L. Bellot Rubio\inst{3}
          }

   \institute{INAF - Institute for Space Astrophysics and Planetology (IAPS),
              Via del Fosso del Cavaliere 100, 00133 Roma, Italy\\
              $^*$ Current address: Istituto Nazionale di Geofisica e Vulcanologia,
              Via di Vigna Murata 605, 00143 Roma, Italy\\  
              \email{fabio.giannattasio@ingv.it}
         \and
             Department of Physics, University of Rome Tor Vergata
             Via della Ricerca Scientifica 1, 00133 Roma, Italy\\
         \and
         	 Instituto de Astrof\'isica de Andaluc\'ia (CSIC), 
             Apdo. de Correos 3004, E-18080 Granada, Spain
             }

   \date{}

 
  \abstract
   {Turbulent convection efficiently transports energy up to the solar photosphere, but its multi-scale nature and dynamic properties are still not fully understood. Several works in the literature have investigated the emergence of patterns of convective and magnetic nature in the quiet Sun at spatial and temporal scales from granular to global.}
   {To shed light on the scales of organisation at which turbulent convection operates, and its relationship with the magnetic flux therein, we studied characteristic spatial and temporal scales of magnetic features in the quiet Sun.}
   {Thanks to an unprecedented data set entirely enclosing a supergranule, occurrence and persistence analysis of magnetogram time series were used to detect spatial and long-lived temporal correlations in the quiet Sun and to investigate their nature.}
   {A relation between occurrence and persistence representative for the quiet Sun was found. In particular, highly recurrent and persistent patterns were detected especially in the boundary of the supergranular cell. These are due to moving magnetic elements undergoing motion that behaves like a random walk together with longer decorrelations ($\sim2$ h) with respect to regions inside the supergranule. In the vertices of the supegranular cell the maximum observed occurrence is not associated with the maximum persistence, suggesting that there are different dynamic regimes affecting the magnetic elements.}
   {}

   \authorrunning{F.Giannattasio et al.}
   \titlerunning{Occurrence and persistence of magnetic elements in the quiet Sun}
   \maketitle
%

\section{Introduction}

Turbulent convection is the mechanism that efficiently transports energy through the outermost 30\% of the solar radius, and is responsible for the formation of space- and time-coherent structures observed at scales from granular \citep[see, e.g., ][and references therein]{1999A&A...344L..29B, 1999A&A...344L..33C, 2002A&A...381..253B, 2004SoPh..221...33B, 2004A&A...428.1007D, 2006A&A...451.1081N, 2007ApJ...666L.137C} to mesogranular \citep[see, e.g., ][]{1980PhDT.........3N, 1998A&A...330.1136R, 2005ApJ...632..677B, 2011ApJ...727L..30Y, 2013SoPh..282..379B} and supergranular \citep[see, e.g., ][]{1956MNRAS.116...38H, 1964ApJ...140.1120S, 2004SoPh..221...33B, 2004SoPh..221...23D, 2008ApJ...684.1469D, 2012ApJ...758L..38O, 2013ApJ...770L..36G, 2014ApJ...788..137G, 2014A&A...569A.121G, 2014ApJ...797...49G, 2014A&A...568A.102B, 2014A&A...561L...6S, 2016ApJ...820...35G}.
Actually, the photospheric features observed on the Sun are the result of the action of turbulent convection and its complex non-linear interaction with the magnetic fields amplified and emerging from the convective region.
This interaction deeply affects the solar activity, and is often invoked to be the basis of those physical processes that trigger the energy transfer in and to the upper layers of the solar atmosphere, such as magnetic reconnections \citep[][]{1957JGR....62..509P, 1988ApJ...330..474P}, which may originate nanoflares \citep[see, e.g., ][]{2006ApJ...652.1734V}, and the excitation of acoustic \citep[see, e.g.,][]{2006ApJ...648L.151J, 2014CEAB...38...53S} and magnetohydrodynamic (MHD) waves \citep[see, e.g., ][and references therein]{1947MNRAS.107..211A, 2007Sci...318.1574D, 2007Sci...317.1192T, 2014A&A...569A.102S, 2015A&A...577A..17S, 2017NatSR...743147S, 0004-637X-840-1-19}.

Nevertheless, despite the efforts, the physics of high Rayleigh number ($Ra$) turbulent convection has major open issues and is still being largely debated \citep[see, e.g., ][]{Xiaozhou2012}.
Actually, values higher than $Ra\sim10^{17}$ cannot be reached in laboratory experiments \citep[e.g., ][]{2000Nature...404N}, while for solar convection $Ra\sim10^{19}-10^{24}$ are expected \citep{2012PNAS..10911928H}.
The inherent difficulty in describing a complex system like the magnetised and highly turbulent photospheric plasma, means that it is still not possible to formulate a complete and extensive theory of solar convection on all scales, from sub-granular to global.
To address this problem, three different approaches have been proposed in the last decades: i) MHD simulations, ii) tracking of small-scale magnetic fields (magnetic elements), and iii) statistical study of the emerging patterns on different space and time scales.

Magnetohydrodynamic simulations have been widely used in the last twenty years to reproduce all the features observed in the solar photosphere, such as granulation and the emerging of both small- and large-scale magnetic fields \citep[see, e.g., ][]{1997ASSL..225...79N, 1998ApJ...499..914S, 2001ApJ...546..585S, 2003ApJ...588.1183C, 2005A&A...429..335V, 2009Sci...325..171R, 2012A&A...539A.121B, 2015A&A...574A..28D}.
As a result, MHD simulations were successful in mimicking accurately limited regions, in volume and time, of the convection zone.
However, despite their capability to match the observations in a very realistic way, at present they cannot simulate simultaneously the wide range of spatial and temporal scales involved in the solar photosphere, from sub-granular to global, due to intrinsic limitations of currently available computing power.

Some inherent aspects of turbulent convection related to advection/diffusion, magnetic flux concentration and organisation on different spatial and temporal scales can be alternatively investigated by tracking bright features (possibly related to magnetic elements) in G-band images \citep[see, e.g., ][]{1998ApJ...506..439B, 1998ApJ...509..918C, 1999ApJ...521..844C, 2001PhRvL..86.5894L, 2010ApJ...715L..26S, 2011ApJ...743..133A, 2012ApJ...759L..17L, 2015RAA....15..569Y, 2015ApJ...810..2Y} or magnetic elements in magnetograms \citep[see, e.g., ][]{1988SoPh..116....1W, 1999ApJ...511..932H, 2011A&A...531L...9M, 2013ApJ...770L..36G, 2014ApJ...788..137G, 2014A&A...569A.121G, 2014A&A...566A..99K, Caroli2015, 2016A&A...590A.121R, 2016arXiv160502533I}, and possibly comparing the results with those expected from simulations obtained with simplified advection/diffusion processes \citep[see, e.g., ][]{2015A&A...576A..47D}.
The accuracy of this approach is based on the hypothesis that magnetic elements are passively transported across the solar photosphere by the plasma flow, so that their motion reveals the physical properties of the underlying velocity field.

The statistical analysis of long-time duration data can allow us to probe the structure of photospheric and sub-photospheric plasma flows and magnetic field distribution at different spatial and temporal scales.
This can be approached, for example, via the investigation of emerging correlations and the detection of coherent patterns \citep[see, e.g., ][]{2002A&A...382L...5G, 2002A&A...392L..13R, 2006SoPh..239...93G, 2008SoPh..249..307B}.

By following approach iii), in this work we investigate spatial and temporal correlations of the photospheric magnetic field in the quiet Sun taking advantage of an unprecedented magnetogram time series acquired on 2 November 2010 with the SOT telescope on board Hinode \citep{2007SoPh..243....3K, 2008SoPh..249..167T} and targeted at the disc centre.
The paper is organised as follows: in \S2 we present the data set used and describe the methods applied to perform the analyses; in \S3 we show the results obtained; in \S4 we discuss the results in light of the existing literature; and in \S5 we summarise the main aspects pointed out in this work and draw our conclusions.

\section{Observation and data analysis}
\subsection{The 24-hour Hinode data set}
The data set used in this work was described exhaustively in \citet{Milan}, and previously analysed by \citet{2013ApJ...770L..36G, 2014ApJ...788..137G, 2014A&A...569A.121G} and \citet{Caroli2015} to perform their studies on the diffusion by turbulent convection of small-scale magnetic fields in the quiet Sun over a wide range of spatial and temporal scales, from granular to supergranular.
It consists of a series of $959$ magnetograms acquired by Hinode SOT \citep{2007SoPh..243....3K, 2008SoPh..249..167T} with a field of view (FoV) of $\sim50$ Mm, a spatial resolution of $\sim0".3$, and a noise level - computed as the rms value in a sub-field of view (sub-FoV) free of magnetic signal convolved with a $3\times3$ spatial kernel - of $\sigma_N\simeq4$ G for single magnetograms. 
The series was filtered for oscillations at $3.3$ mHz \citep{Milan} in order to remove the effect of acoustic oscillations.
We note that our observations are also free from any seeing effect.
Particularly interesting, the large FoV contains an entire supergranule.
This, together with the high spatial resolution and the absence of seeing, makes the data set very suitable for studying the evolution and structure of the quiet Sun as a consequence of the plasma dynamics induced by turbulent convection.
Moreover, the series of data, acquired on 2 November 2010, spans 24 hours without interruption, with a cadence of $90$ s.
\begin{figure}[ht!]
	\centering
	\includegraphics[width=10cm]{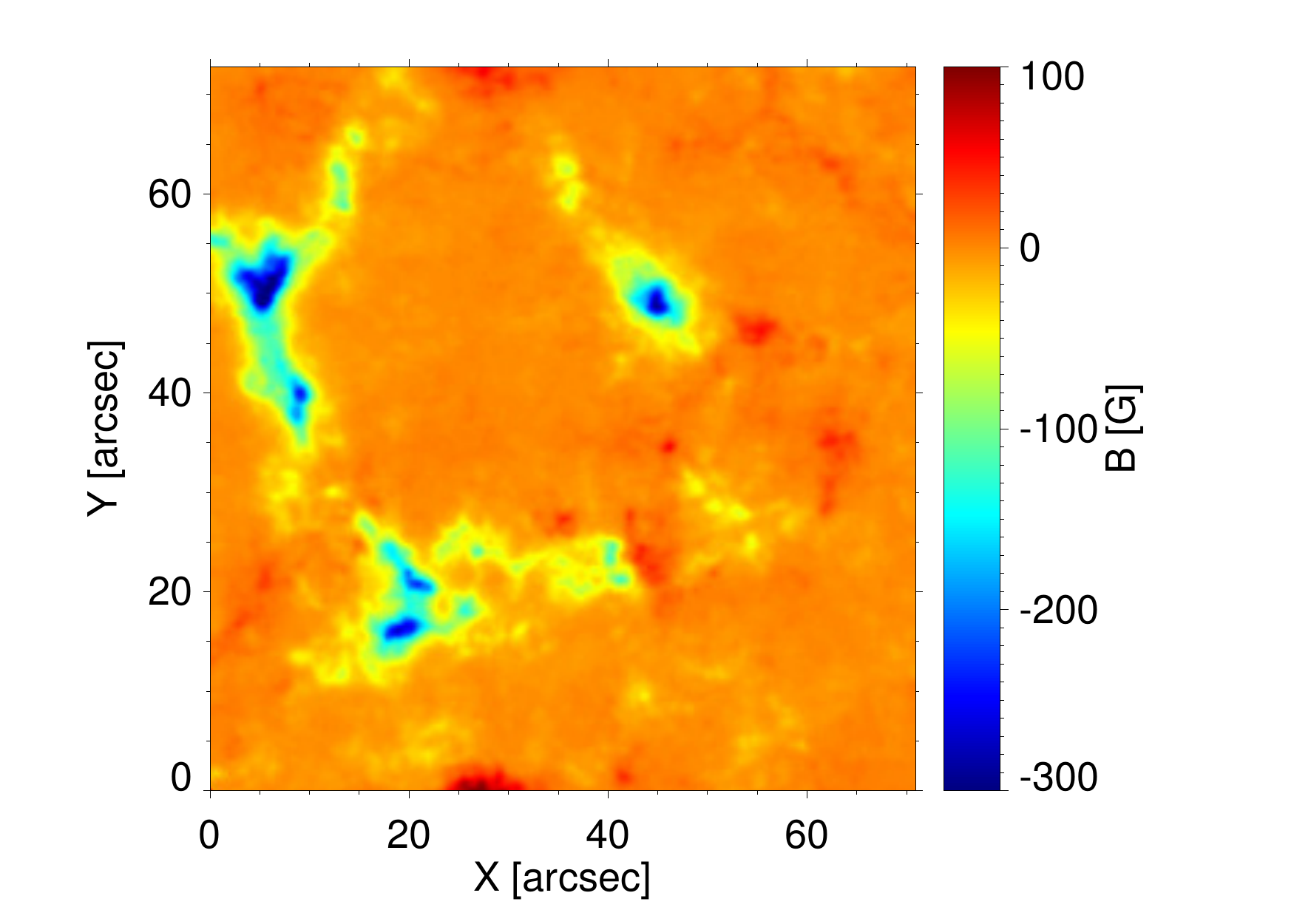}
	\caption{Mean magnetogram of the FoV averaged over $\sim24$ hours, the whole time range of the series.}
    \label{Fig.mean_B_Frame}
\end{figure}


In Fig.\ref{Fig.mean_B_Frame} we show the signed mean magnetogram averaged over $\sim24$ hours.
The boundaries of the supergranular cell are clearly visible, and also outlined by the horizontal velocity maps obtained via the application of the {\it Fast Local Correlation Technique} (FLCT) described in \citet{2008ASPC..383..373F} and shown in the bottom panels of Figure \ref{Fig.maps} \citep[see also Figures 1a and 1b in][]{2014ApJ...788..137G}.
The horizontal velocity field is mainly radial and directed from the central regions to the boundaries of the supergranular cell, with the velocity strength ranging from $\sim0.1$ to $\sim0.6$ kms$^{-1}$ moving outward.

\subsection{Occurrence and persistence of magnetic features}
In what follows, we indicate with $\Phi$ the unsigned magnetic flux density strength.

\subsubsection{Occurrence}
For any pixel $(x,y)$ of the FoV, we defined the {\it occurrence}, $R_{xy}$, as the number of frames in which that pixel hosted a magnetic feature with strength $\Phi_{xy}$ above a magnetic threshold $\Phi_T$, i.e. the number of times that the condition $\Phi_{xy}>\Phi_T$ is fulfilled.
According to this definition, $R_{xy}$ is a measure of the tendency of a specific site to host magnetic features.

\begin{figure}[h!]
	\centering
	\includegraphics[width=8cm]{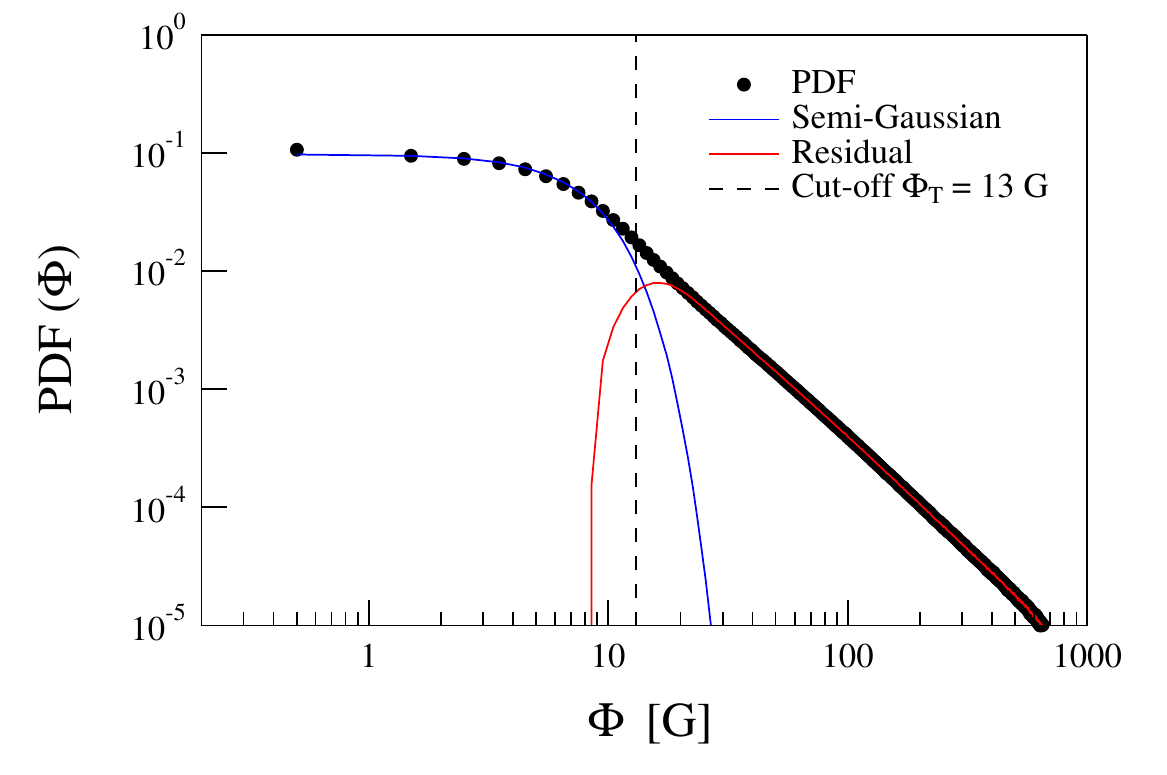}
	\caption{Flux strength probability distribution function, $PDF(\Phi)$, obtained by considering all the frames in the magnetogram time series (black dots).The blue line corresponds to a semi-Gaussian fitting model, the red line to the residual. The threshold value of $\Phi_T=13$ G is marked by a vertical dashed line.}
    \label{Fig.PDF_B}
\end{figure}

The occurrence of magnetic features was investigated by performing a numerical binarisation of each image of the time series.
This was obtained by setting to unity all pixels with a magnetic flux strength above $\Phi_T=13$ G, zero elsewhere.
The threshold $\Phi_T$ corresponds to two values.  
The first is the magnetic flux strength at which the flux strength probability distribution function (PDF) departs from a semi-Gaussian shape with standard deviation $s$, namely $g(\Phi,s)$, representative of the noise
\begin{equation}
g(\Phi,s)=\frac{\sqrt{2}}{s\sqrt{\pi}}exp\left(-\frac{\Phi^2}{2s^2}\right) \qquad \Phi\ge0.
\label{Eq.Semi-G}
\end{equation}
This threshold is defined as \citep{CeD2002}
\begin{equation}
\Phi_T=\frac{\int_0^\infty\Phi g(\Phi,s) r(\Phi,s)d\Phi}{\int_0^\infty g(\Phi,s) r(\Phi,s)d\Phi},
\end{equation}
where $r(\Phi,s)$ is the residual distribution obtained as the difference between the observed PDF and $g(\Phi,s)$ (see Figure \ref{Fig.PDF_B}).
The second is the commonly used threshold of three times the magnetogram noise, $3\cdot\sigma_N$. \\
Due to both i) and ii) the threshold $\Phi_T$ represents a robust value (with a probability over $99.7\%$) to discern between magnetic features and noise.

The sum of binarised images normalised to their total number and expressed in percentage provided the {\it normalised occurrence}, $O_{xy}$, which can be defined pixel-by-pixel as
\begin{equation}
O_{xy}=\frac{R_{xy}}{N}\cdot100,
\end{equation}
where $N$ is the number of magnetograms included in the time series.
We note that with respect to $R_{xy}$, $O_{xy}$ provides a normalised value in the range $[0,1]$ for all possible magnetogram series. 
For practical purposes, we refer to occurrence as the quantity $O_{xy}$ throughout the text.

\subsubsection{Persistence}
For any pixel in the FoV we defined the {\it persistence} as the time range, $t_D$, over which the magnetic flux is still autocorrelated.
According to this definition, $t_D$ represents the characteristic time range during which {\it the same} magnetic features linger in a specific location.
In order to study the pixel-by-pixel persistence in the magnetogram time series and estimate $t_D$ we proceeded as follows.
For any time step and correlation time $(t, t_C)$ in the range $0<(t, t_C)\le (N-1)\Delta t$, being $\Delta t$ the time cadence of observations, we computed the correlation Pearson coefficient in order to obtain the auto correlation function \citep[ACF, see, e.g., ][]{1986nras.book.....P}
\begin{equation}
ACF_{xy}=\frac{cov(\Phi_{xy}(t),\Phi_{xy}(t+t_C))}{\sigma^2_{xy}},
\label{Eq.rho}
\end{equation}
being $\sigma^2_{xy}$ the variance of $\Phi_{xy}(t)$, namely
\begin{equation}
\sigma^2_{xy}=\overline{(\Phi_{xy}(t)-\bar{\Phi}_{xy})^2},
\label{Eq.sigmas}
\end{equation}
and
\begin{equation}
cov(\Phi_{xy}(t),\Phi_{xy}(t+t_C))=\left\langle (\Phi_{xy}(t)-\bar{\Phi}_{xy})(\Phi_{xy}(t+t_C)-\bar{\Phi}_{xy})\right\rangle,
\label{Eq.cov}
\end{equation}
where in equations \ref{Eq.sigmas} and \ref{Eq.cov} the overbar stands for the mean value and the brackets $\langle...\rangle$ correspond to an ensemble average over all the pairs of the time series $\Phi_{xy}(t)$ separated by $t_C$.

The ACFs computed via equations \ref{Eq.rho}-\ref{Eq.cov} peak at $t_C=0$ (where ACF is unity) and then decrease to zero in a characteristic time which is longer where the magnetic flux change is slower.
For each pixel of the FoV, we estimated this decorrelation time, $t_D$, as the time at which the ACF decreases from unity to the quantity $1/e$.
We note that $t_D$ is univocally defined, as there is only one time at which $ACF=1/e$.
In fact, for $t>t_D$ the ACF decreases and oscillates around zero with amplitude well below $1/e$.

\section{Results}
\label{Sec.Results} 
\begin{figure*}[ht!]
	\centering
	\includegraphics[width=9cm]{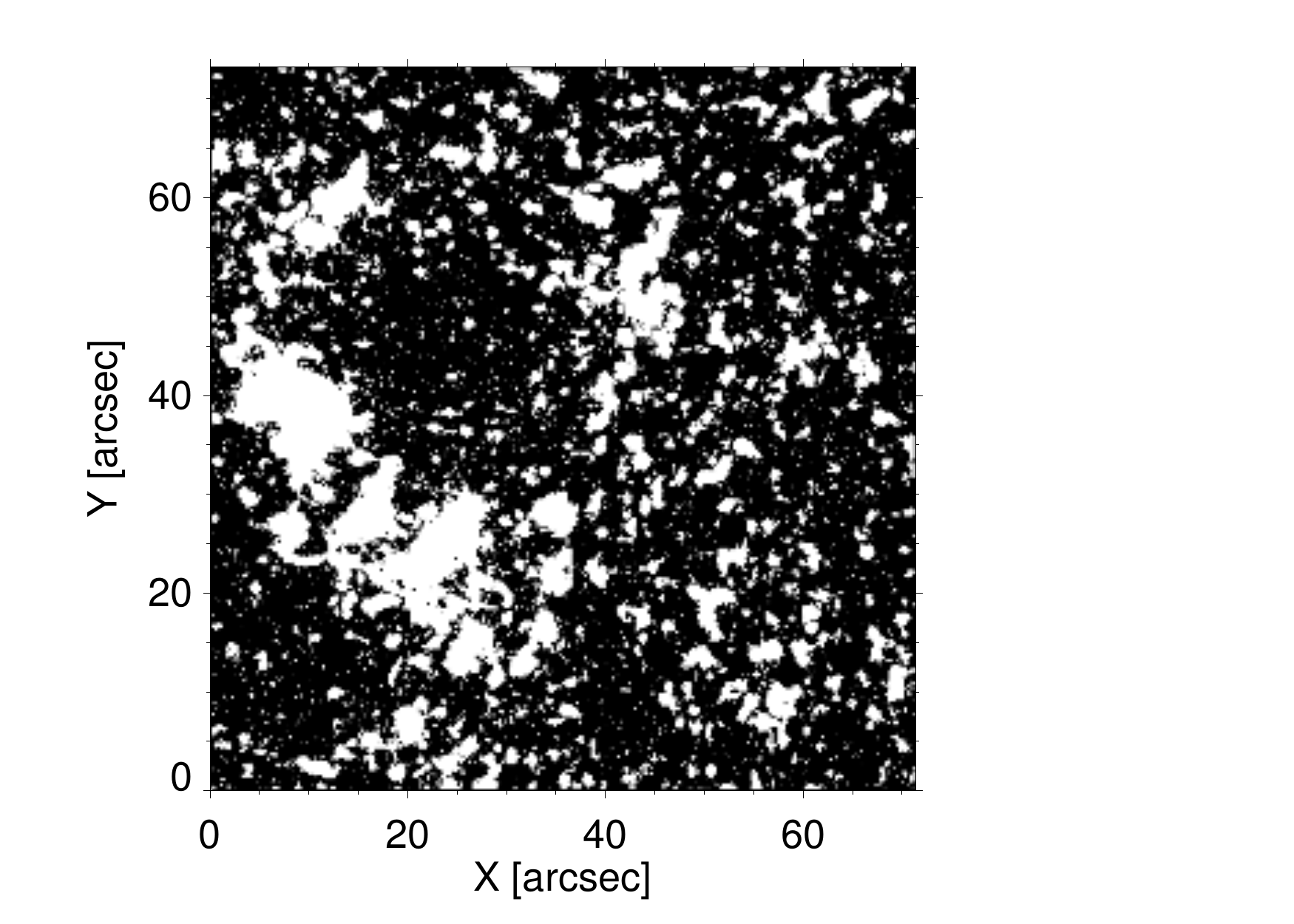}
	\includegraphics[width=9cm]{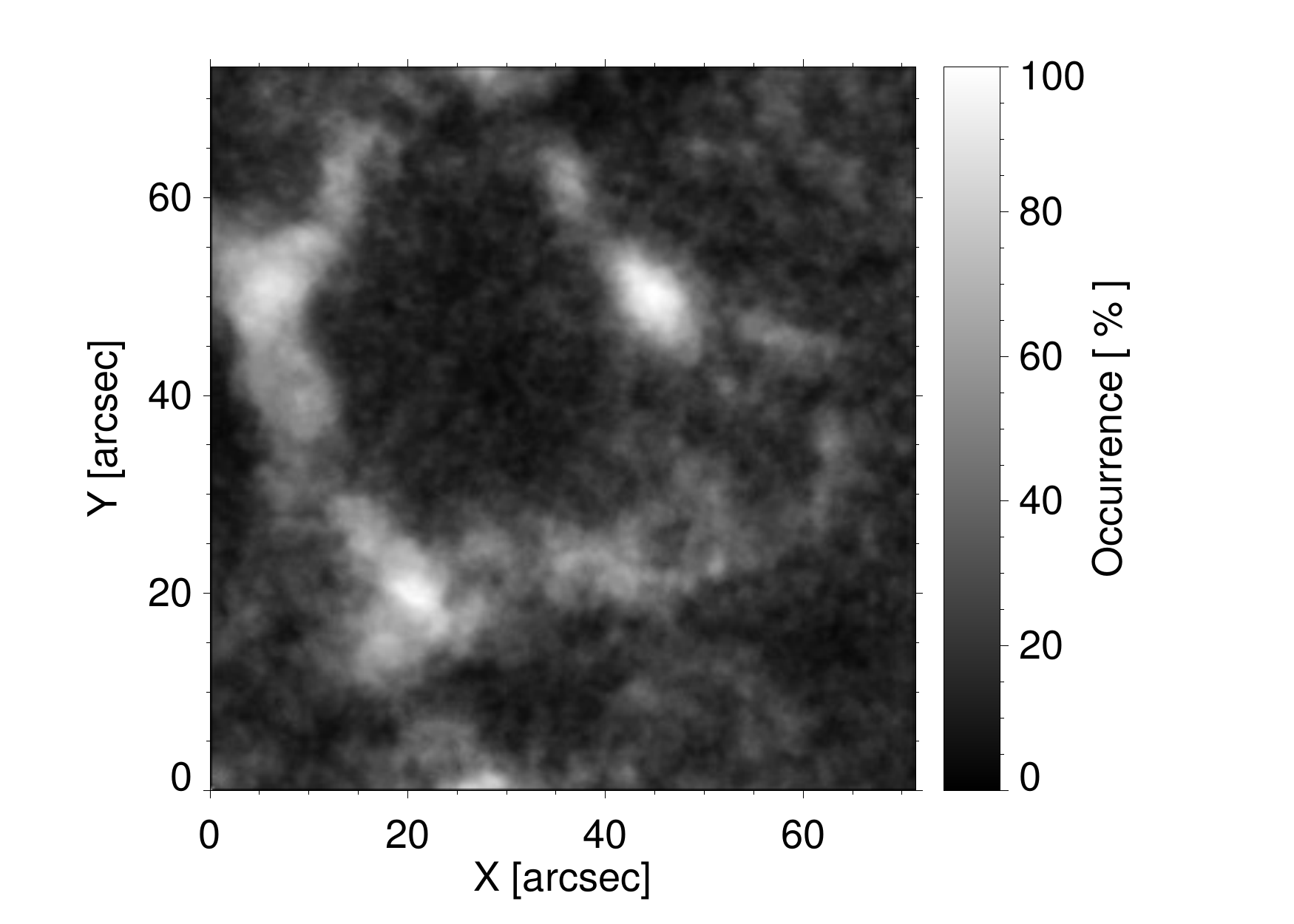}
	\includegraphics[width=9cm]{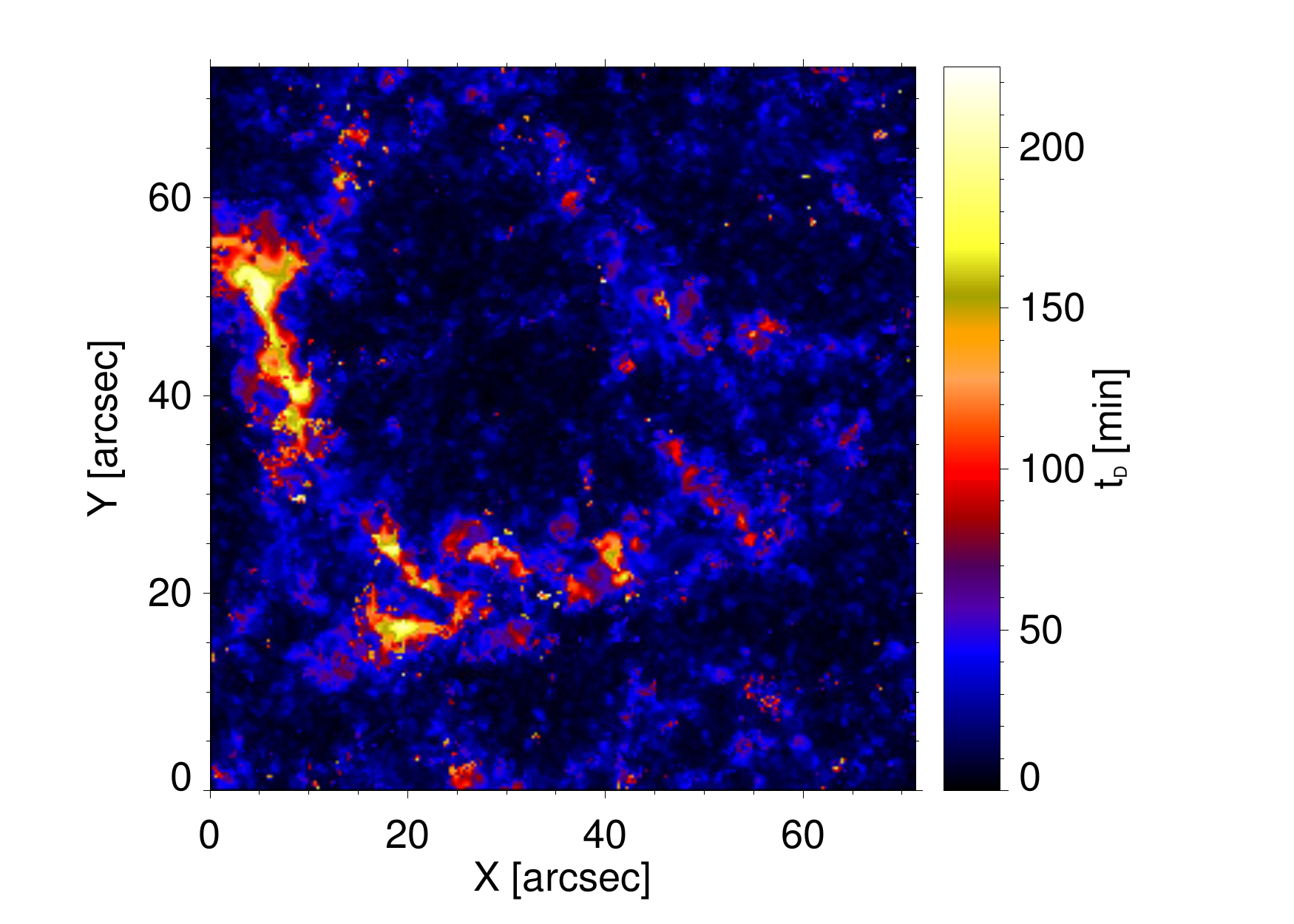}
	\includegraphics[width=9cm]{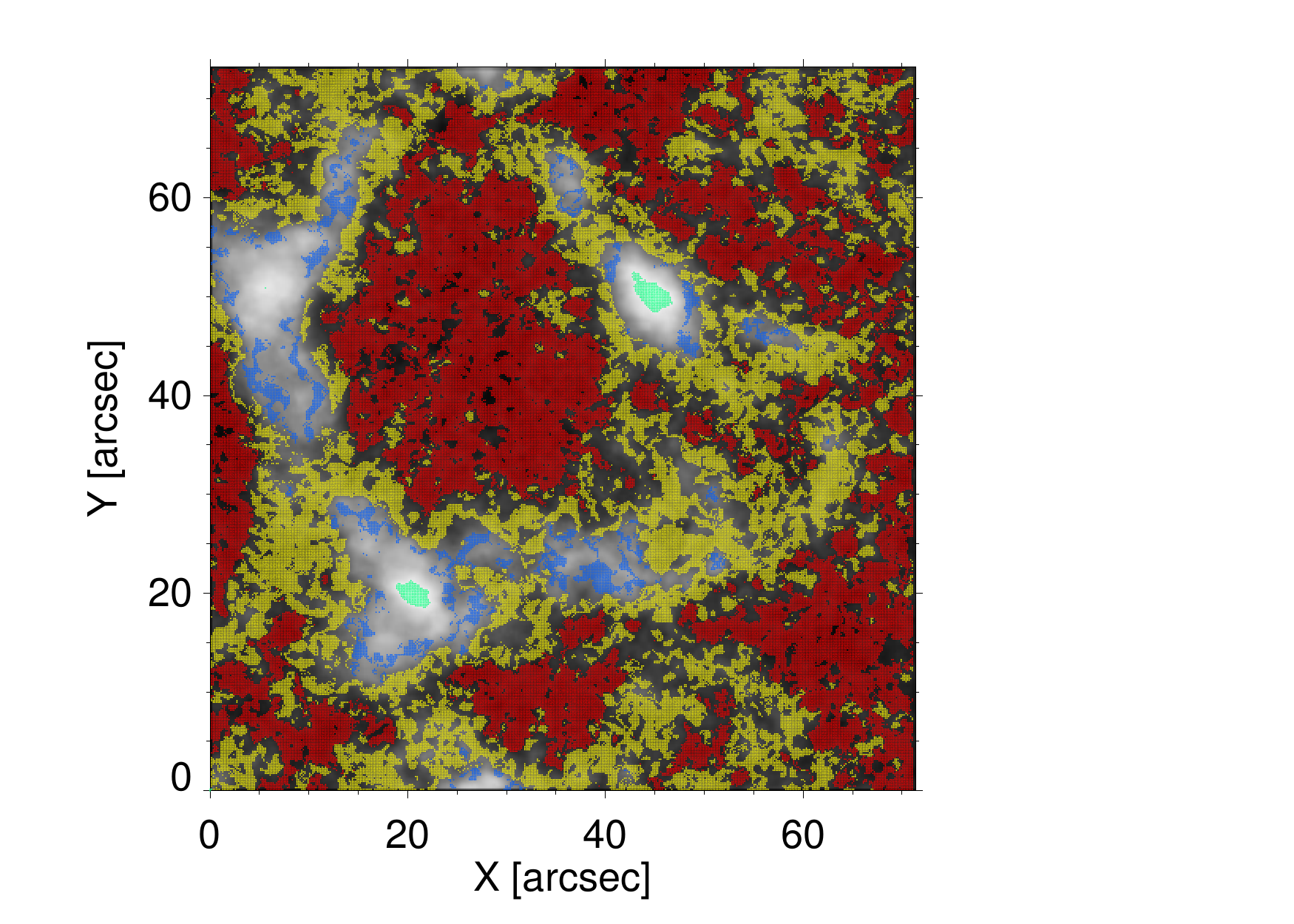}
	\includegraphics[width=9cm]{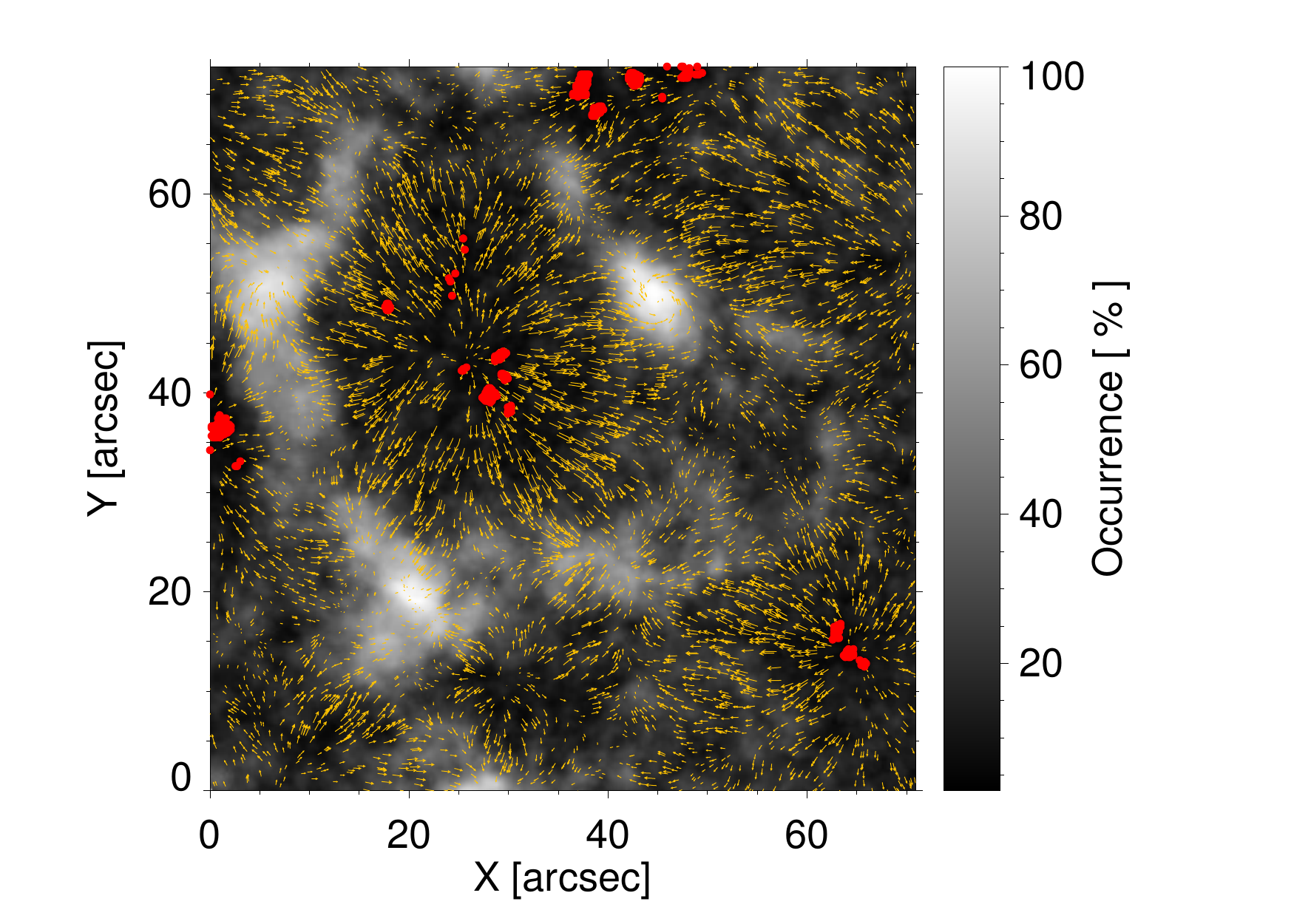}
	\includegraphics[width=9cm]{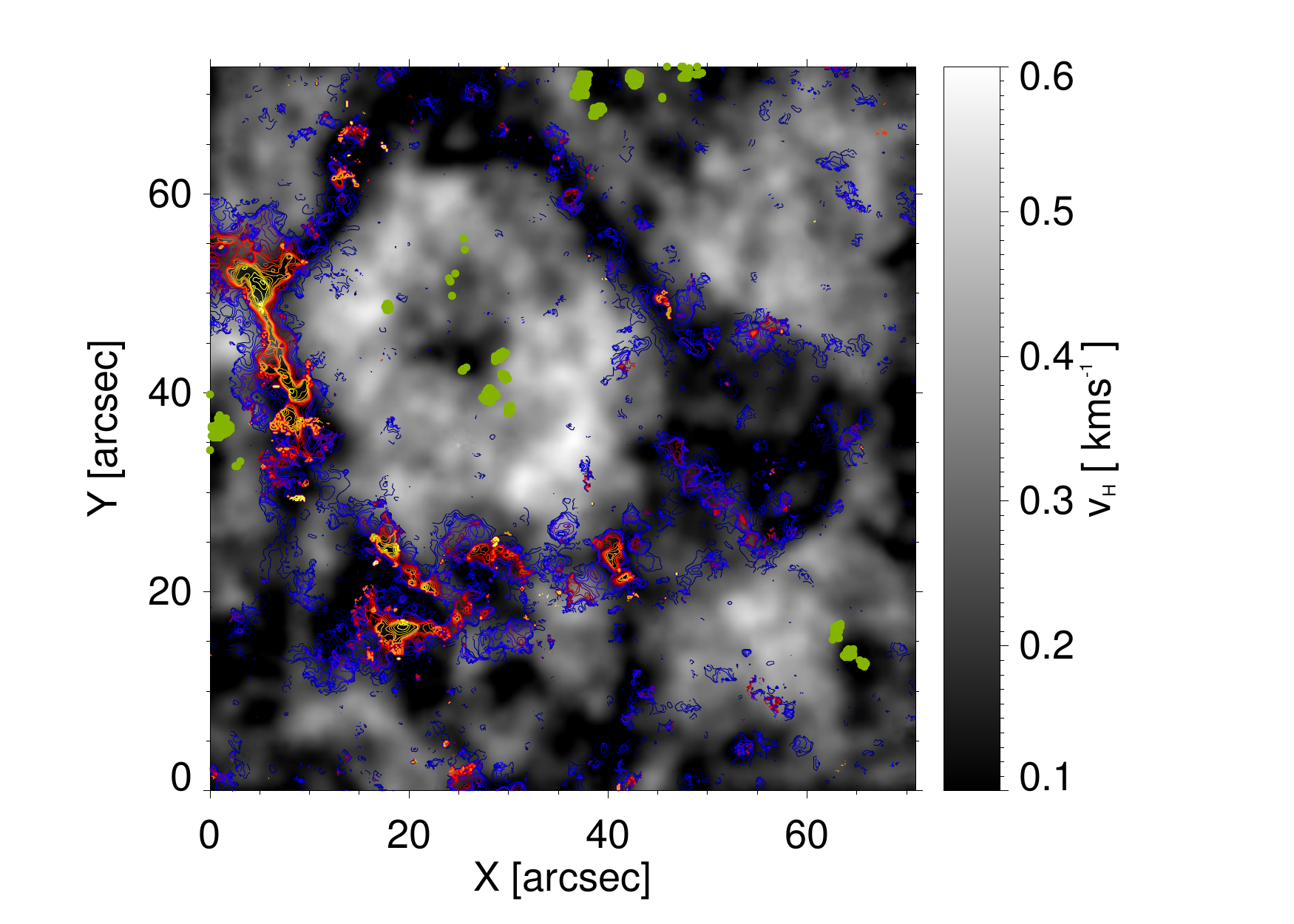}
    \caption{{\it Top left}: Binarisation map of the first magnetogram of the series, i.e. acquired at 08:00:42 UT. Pixels in white are those with magnetic flux density $\Phi\ge\Phi_T=13$ G. {\it Top right}: Occurrence map of the FoV. {\it Middle left}: Map of decorrelation times. {\it Middle right}: Map of the different regions identified in Figure \ref{Fig.Regions} in red, yellow, blue, and green (see text) superimposed on the occurrence map. {\it Bottom left}: Horizontal velocity field (gold arrows) computed with FLCT \citep{2008ASPC..383..373F} superimposed to the occurrence map. The red filled circles mark the locations with the lowest occurrences ($\le5\%$). {\it Bottom right}: Horizontal velocity strength map (grey scale). The decorrelation times map for $t_D>20$ min is superimposed with the same color code as in the middle left panel. The green filled circles mark the locations with the lowest occurrences ($\le5\%$).}
    \label{Fig.maps}
\end{figure*}

\begin{figure}[ht!]
	\centering
	\includegraphics[width=8cm]{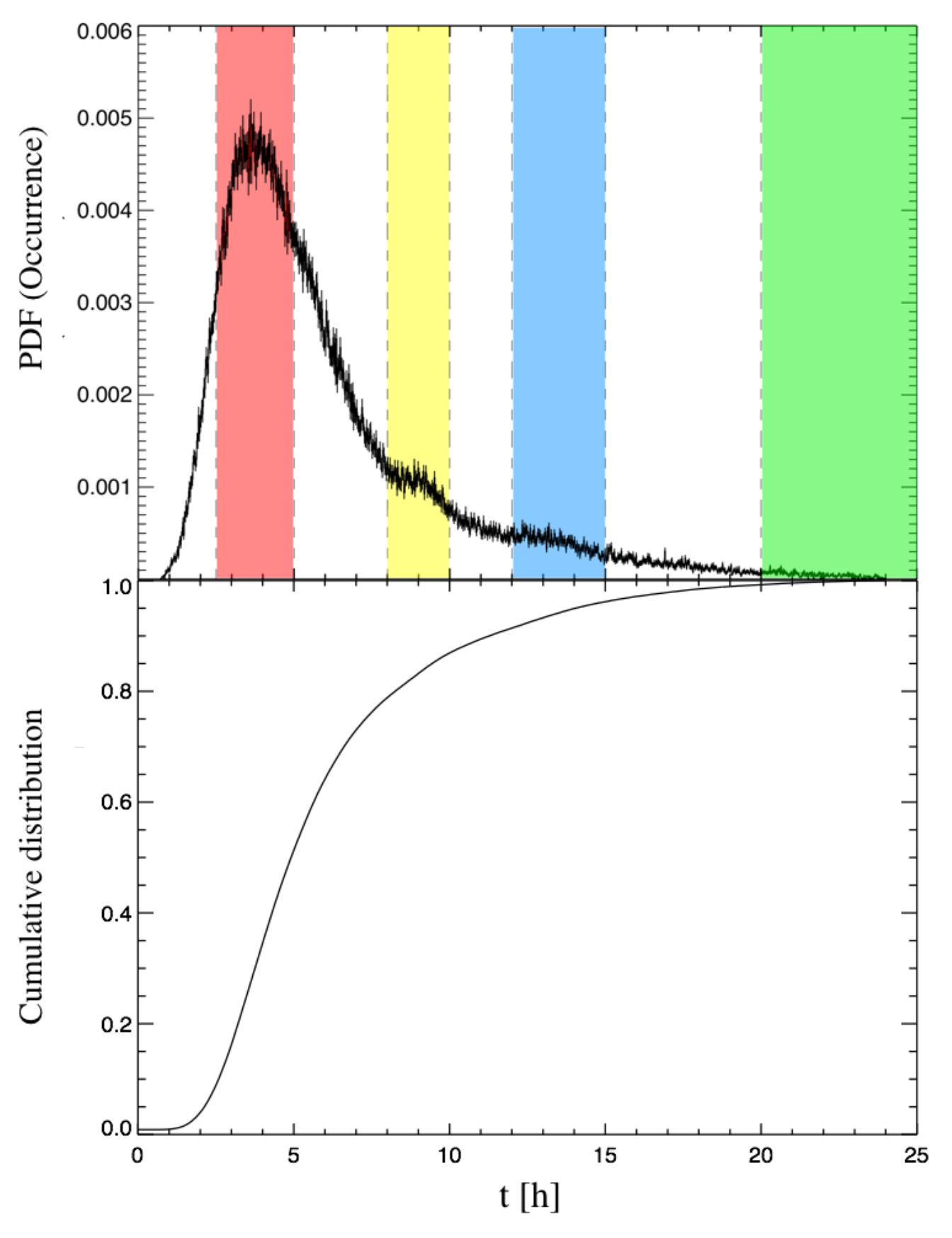}
	\caption{{\it Upper panel}: PDF of occurrence. The different colours highlight different features in the distribution, corresponding to the main peak (red), two minor peaks (yellow and blue), and the most 'extreme' events (green). {\it Lower panel}: Cumulative distribution of occurrence.}
    \label{Fig.PDF_Occurrence}
\end{figure}

In the top left panel of Figure \ref{Fig.maps} we show the results of binarisation applied to the first magnetogram of the series.
The white areas in the figure correspond to regions with magnetic flux density $\Phi\ge\Phi_T$. 
The largest areas that are correlated are recognisable as the boundaries of the supegranular cell.
The top right panel of the same figure shows the occurrence map obtained by considering the whole data set.
An occurrence greater than $95\%$ is found in just $266$ pixels (corresponding to $\sim0.1\%$ of the FoV), each located at a vertex of the supergranular cell.
In the supergranular boundary, the occurrence always exceeds $\sim35-40\%$, while values of a small percent are found near the centre of the supergranular cell and in those regions of the FoV that probably correspond to the innermost parts of the adjacent supergranules \citep[bottom panels of Figure \ref{Fig.maps} and][]{Milan, 2016ApJ...820...35G}. 
In the upper panel of Figure \ref{Fig.PDF_Occurrence} we show the PDF of occurrence, which this time was expressed in hours instead of percentage in order to easily identify possible characteristic time scales.
A peak is observed between $2.5$ and $5$ hours (red region in the figure), where the probability is $\simeq0.5\%$.
Then the PDF smoothly decreases to $\simeq8$ hours.
A small secondary peak is observed between $8$ and $10$ hours (yellow region in the figure), where the probability increases to $\simeq0.12\%$.
After a slight decrease a last minor peak is observed between $12$ and $15$ hours (blue region in the figure), where the probability increases to $\simeq0.06\%$.
Then the PDF slowly converges to zero.
The region between $20$ and $24$ hours is in green in the figure.

The lower panel of Figure \ref{Fig.PDF_Occurrence} shows the cumulative distribution of occurrence.
About $50\%$ of the pixels in the FoV have occurrences below $\simeq5$ hours, thus lying on the left of the main peak, while only $\simeq0.8\%$ of pixels have occurrences above $20$ hours.

The occurrence map gives statistical information on the sites hosting magnetic elements with $\Phi\ge\Phi_T$, and indicates where magnetic elements are likely to be found in the FoV over the whole time range spanned in our observations.
Complementary information is provided by the characteristic time over which each pixel in the FoV hosts correlated magnetic elements, i.e. the persistence of magnetic elements at any location.
In the middle left panel of Figure \ref{Fig.maps} we show the map of pixel-by-pixel decorrelation times, $t_D$, for the magnetogram time series, which was computed as the time at which the ACF drops by a factor $1/e$.
As we can see, $t_D$ spans the range between $\simeq30$ and $\simeq240$ minutes in the supergranular boundary.
Inside the supergranule, the decorrelation time map shows dark regions where $t_D$ is a few minutes at most, together with enhanced patterns where $t_D$ is typically $\simeq30-50$ minutes, and a few peaks characterised by a $t_D\simeq80$ minutes.
In the remaining part of the FoV, alternating bright and dark structures are observed, with values of $t_D$ ranging from zero up to $\sim100$ minutes, the latter occurring in correspondence with local maxima in the spatial distribution of magnetic flux (the most evident is at location [$X=26$, $Y=3$ arcsec] in the FoV).

\begin{figure*}[ht!]
	\centering
	\includegraphics[width=12cm]{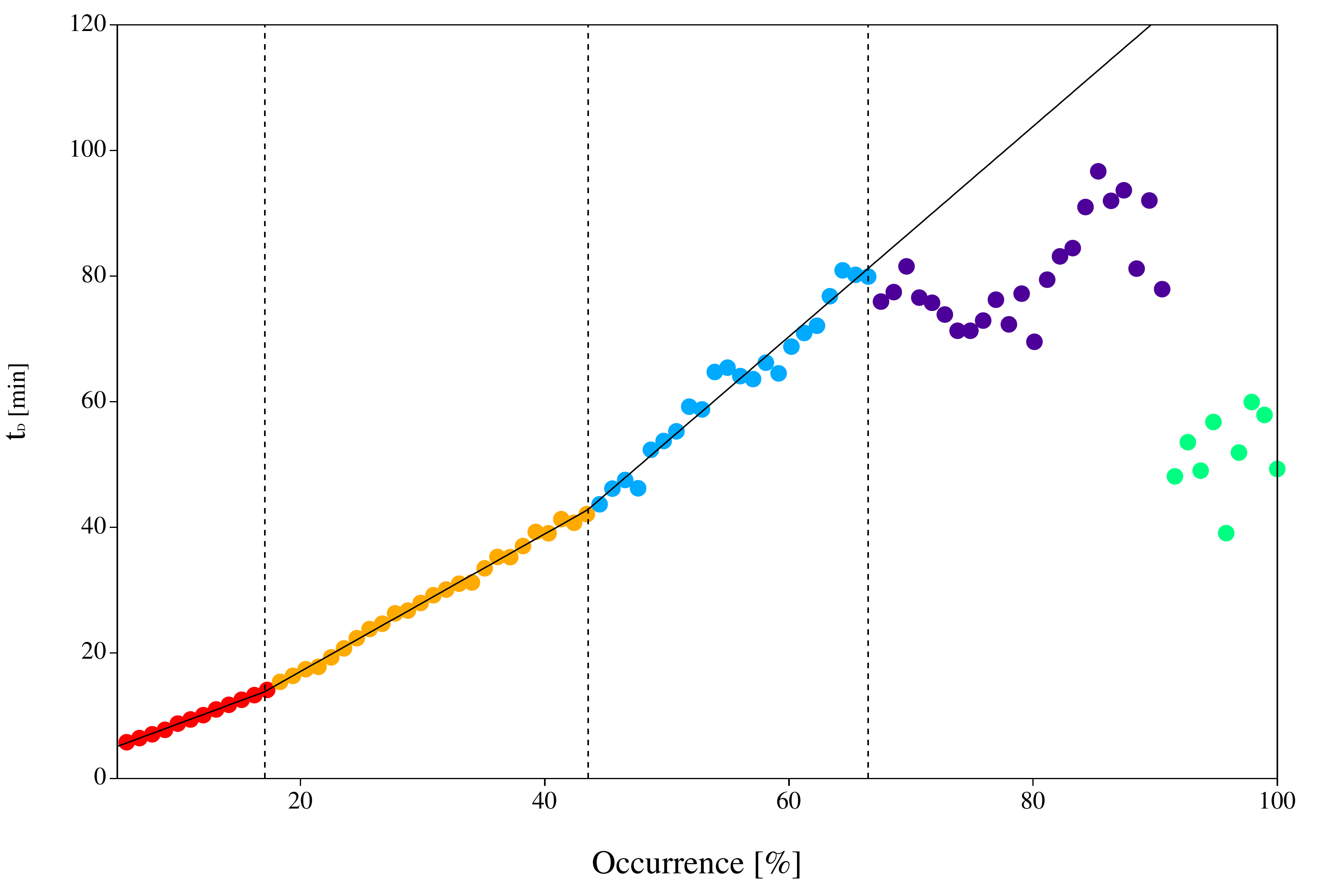}
	\begin{picture}(0,0)
	\put(-300,124){\includegraphics[height=3.4cm,width=5.3cm]{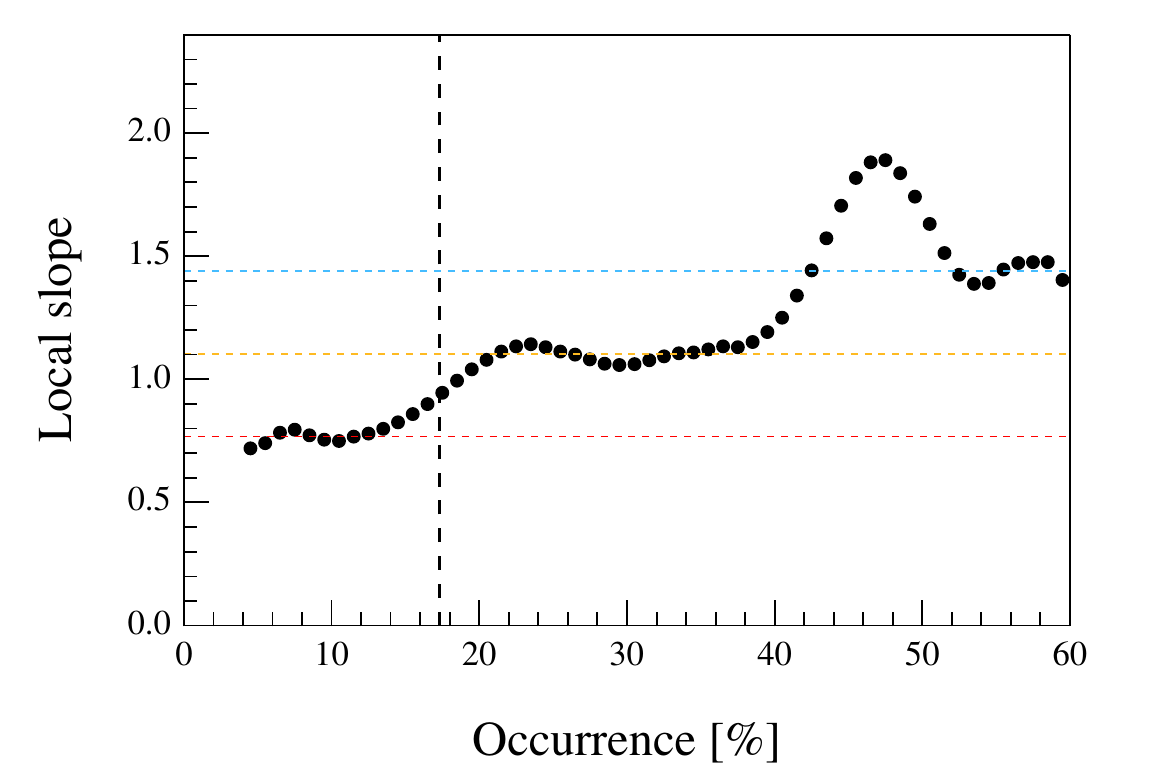}}
	\end{picture}
	\caption{Scatter plot between occurrence (in percentage) and the average decorrelation time (in minutes) in bins of occurrence $\sim1\%$ wide (corresponding to $10$ frames). The different colours and the vertical dashed lines separate different regimes in the plot. The solid line corresponds to the best fit of data from a polygonal model up to an occurrence of $65\%$. In the top left inset: local slope of data points as a function of occurrence. The different regimes detected correspond to nearly constant slopes of $0.77$ (in red), $1.10$ (in orange), and $1.44$ (in blue).}
    \label{Fig.Regions}
\end{figure*}
The top right and middle left panels of Figure \ref{Fig.maps} show similar features underlining the boundary of the supergranular cell.
Longer $t_D$ are necessarily associated with higher occurrences, conversely higher occurrences do not imply longer $t_D$.  
We investigated the relation between occurrence and persistence in the FoV by means of conditioned statistics.
In Figure \ref{Fig.Regions} we show the scatter plot between occurrence (in percentage) and the average $t_D$ (in minutes) in bins of occurrence $\sim1\%$-wide (corresponding to $10$ frames).
Two different guess functions were used to best fit the data, a fourth-order polynomial function, and a polygonal piecewise linear function.
The most suitable number of change points of the polygonal function was evaluated via the local slope of $t_D$ as a function of the occurrence, which was computed by performing a linear fit of $t_D$ on a nine-point mobile window centred on each occurrence (top left inset in Figure \ref{Fig.Regions}).
The best number of change points we identified is three, and corresponds to the number of ranges in occurrence within which the local slope changes substantially.
An automated procedure based on the minimisation of $\chi^2$ over a mobile window selected the polygonal function (solid black line in Figure \ref{Fig.Regions}) as the best to fit the data.
In fact, we found a reduced $\chi^2=0.19$ for the polygonal function, and $\chi^2=0.44$ for the fourth-order polynomial function, respectively.
The automated procedure also provided the values of the change points at occurrences $17\%$, $43\%$, and $65\%$ (marked by vertical dashed lines in the same figure).
These change points separate different regimes, which are characterised by different local slopes and highlighted with different colours in Figure \ref{Fig.Regions} for the sake of clarity.
It is possible that another regime may emerge at even higher occurrences ($>65\%$), but the lack of statistics there prevented us from suitably estimating the fit parameters.
For occurrences above $60\%$, the computation of the local slope is also heavily affected by statistical effects, and thus is not considered reliable.
For occurrences $\lesssim60\%$, the behaviour of the local slope suggests the presence of different dynamic regimes, characterised by quasi-constant values of the derivative (at least in the occurrence ranges $\leq17\%$ and $17\%-43\%$) and separated by transition regions where the derivative is subjected to steeper variations.
The different regimes correspond to quasi-constant slopes (dashed horizontal lines in the inset) of $0.77$ (in red), $1.10$ (in orange), and possibly $1.44$ (in blue) being the change points located, as expected, in the middle of the transitions and in correspondence with the vertical dashed lines.
We note that if data were best modelled by a single and smoother function instead of a polygonal one, the local slope should increase monotonically and not show the piecewise linear trend visible in the inset of Figure \ref{Fig.Regions}.
This supports the existence of different regimes to which magnetic elements in the quiet Sun are subjected.

A completely separated population of points is shown in green in the same figure.
These points, together with those belonging to the three different regimes found, may be localised in the FoV.
The results are shown in the middle right panel of Figure \ref{Fig.maps}, where the colour-code used is the same as in Figure \ref{Fig.Regions}, and roughly corresponds to the different regions found in the PDF shown in the upper panel of Figure \ref{Fig.PDF_Occurrence}. 
In moving from the inner regions of the supergranule to the boundaries, the relation between occurrence and $t_D$ steepens.
This increase is broken in the vertices of the supergranule, where a decrease in $t_D$ is observed even though the occurrence reaches its maximum values.  

\section{Discussion}
The flux enhancements in the mean magnetogram of Figure \ref{Fig.mean_B_Frame} underline the largest and longest observed scales of organisation of magnetic elements, which obey a dynamical regime that is different in the supergranular boundary with respect to the intranetwork regions \citep{2014ApJ...788..137G}.
Such enhancements may be due to recurrent and/or persistent magnetic elements, rather than magnetic elements with very high magnetic strength and short lifetime.
In fact, only a small fraction of the FoV has been found to be covered by magnetic flux strengths above the equipartition value, and in any case such strong flux locations host the longest-lived magnetic elements \citep{2013ApJ...770L..36G}.

Further constraints on the magnetic flux evolution up to supergranular scales may be given in terms of the statistical analysis of occurrence and persistence patterns.
In particular, high occurrence values point out the presence of preferred sites hosting magnetic elements, while a high persistence points out the tendency of a site to host the {\it same} magnetic elements, or more precisely the same magnetic flux amount.   
As we can see from the occurrence map (top right panel of Figure \ref{Fig.maps}), the highest values of occurrence are found in the vertices of the supergranular cell where the strongest magnetic fluxes are observed in correspondence with the longest-lived magnetic elements \citep{2013ApJ...770L..36G}.
On the other hand, the lowest occurrences ($\le5\%$, represented as red and green filled circles in the bottom left and right panels of Figure \ref{Fig.maps}, respectively) are found in the central region of the supergranule and of adjacent supergranules \citep{Milan, 2016ApJ...820...35G} where the horizontal velocity as computed by FLCT \citep[][]{2008ASPC..383..373F} is in the range $0.1-0.25$ kms$^{-1}$.
This provides information on the evolution of magnetic flux in supergranular environments.
In fact, if magnetic flux above the detection threshold $\Phi_T$ emerged preferentially in the central regions of the supergranule and then was advected to the cell's boundaries, a high occurrence at the centre should be observed.
The higher the occurrence, the faster the emergence rate.
Moreover, due to the slow horizontal velocity field, the central regions of the supergranule should also be characterised by longer persistence. 
This is not the case, thus the magnetic flux above the detection threshold should emerge in the outer regions of the supergranular cell, as also argued by e.g. \citet{2014A&A...561L...6S} and \citet{2014ApJ...788..137G}, and be quickly advected to the boundaries by the horizontal velocity field, which in these regions reaches $\simeq0.6$ kms$^{-1}$.  

When looking at an enhanced-contrast version of the occurrence map (not shown here), we notice the presence of quasi-radial white features inside the supergranule that resemble those observed by \citet{2014ApJ...788..137G} in the deep magnetogram shown in Figure 1d; they are separated by regions in which the occurrence is $\lesssim10\%$.
These features indicate that the occurrence of magnetic elements is not uniform even inside the supergranular cell, being enhanced in preferred sites which represent paths statistically more traveled than others by magnetic elements.

The PDF of occurrence, obtained by considering only the pixels in which the magnetic flux strength is $\Phi>\Phi_T$ and shown in the upper panel of Figure \ref{Fig.PDF_Occurrence}, suggests the emergence of a characteristic time scale at $\simeq3-4$ hours, and possibly at $\simeq9$ hours, where the observed peak is still significant. 
We also notice that the occurrence ranges identified in the upper panel of Figure \ref{Fig.PDF_Occurrence} and shown with different colours are unambiguously associated with specific magnetic environments in the FoV.
In fact, when searching for the pixels of the FoV corresponding to those occurrence ranges we obtain a map similar to that shown in the middle right panel of Figure \ref{Fig.maps}.
Thus, the emerging time scales may correspond to the typical occupancy time of a certain region of the supergranule by the magnetic flux.


While a high occurrence indicates that a location frequently hosts magnetic features, no matter if the same or different ones, a high persistence is likely to be due to the slow evolution of magnetic elements or to their permanence in a certain region.
Thus, magnetic elements staying for a long time in the same location should result in high persistence, while more dynamic magnetic elements moving to preferred sites (or along preferred paths) should result in low persistence but high occurrence. 
\citet{2014ApJ...788..137G} found that magnetic elements in the supergranular boundaries remain confined therein due to diffusion rates that are slower than those located inside the supergranule.
According to this, the longer-correlated regions shown in the middle left panel of Figure \ref{Fig.maps} correspond to areas hosting magnetic elements with magnetic flux strong enough to oppose to photospheric advection, allowing them to evolve following a nearly random walk and to remain confined in the same regions for a longer time.
This argument is supported by the bottom right panel of Figure \ref{Fig.maps}, where it is shown that the regions with $t_D>20$ min are located in correspondence with the minimum values of the horizontal velocity field.
Outside the supergranule, the other regions in which $t_D$ is enhanced probably correspond to the boundaries of adjacent supergranules \citep[see, e.g., ][]{2014ApJ...797...49G, 2016ApJ...820...35G}.
The drop in $t_D$ inside the supergranule is probably due to the superdiffusive motion of magnetic elements there \citep{2014ApJ...788..137G}, which are dragged along the locally stronger (up to $\simeq 0.6$ kms$^{-1}$) horizontal velocity fields (bottom panels of Figure \ref{Fig.maps}).
Due to these local superdiffusive flows, magnetic elements tend not to remain in the same regions for long, quickly emptying the magnetic flux from the central regions of the supergranule.
Thus, following the horizontal plasma velocity field, magnetic elements are radially expelled from the centre to the outer regions, possibly dragged by trees of fragmenting granules \citep[TFGs, ][]{2016A&A...590A.121R}.
We notice two other features present in the decorrelation map.
First, we see large inhomogeneities in the values of $t_D$ over the boundary of the supergranule, which is the signature of the importance of the properties of local environment.
Such an asymmetry may be the result of the dynamic competition between adjacent supergranules keeping long-correlated magnetic flux frozen in the border area in the left-side network rather than in the right side.
This hypothesis cannot be actually verified, as the horizontal velocity field all around the supergranule cannot be reliably computed everywhere. 
Another possible explanation lies in the presence of unipolar (negative) magnetic flux in the left-side network, and mixed polarity magnetic flux in the right-side network (see Figure \ref{Fig.mean_B_Frame}).
The polarity mixing may result in flux reconfigurations in the right-side network, originating shorter decorrelation times than in the left-side network.
Second, spotty and filamentary features are observed inside the supergranule.
As these features are also present in the occurrence map, and due to the superdiffusive regime observed inside the supergranule, they may be the signature of long-correlated paths along which coherent magnetic fluxes continuously emerge and travel as bricks of the network.

We investigated the relation between occurrence and persistence by plotting the mean $t_D$ within each occurrence bin (Figure \ref{Fig.Regions}).
The trend monotonically increases up to an occurrence $\simeq65\%$, then features appear that are probably due to poor statistics.
Moreover, the higher the occurrence, the steeper the increase in $t_D$.
The best fits of points, based on the minimisation of $\chi^2$ together with the transitions observed in the derivative, provided us with three different regimes of increasing $t_D$.
These different regimes are not randomly located across the FoV, but correspond to coherent regions with different physical conditions, as shown in the middle right panel of Figure \ref{Fig.maps}.
This suggests the presence of different dynamic regimes in action in supergranules, depending on the local environment, which may be the manifestation of different magneto-convection regimes characterising the interaction between photospheric plasma and magnetic field in the quiet Sun \citep{2003ApJ...588.1183C}.

When moving from the centre to the boundary of the supergranule the steepening of $t_D$ with the occurrence indicates a higher coherence of magnetic elements, and the locations with higher probability of hosting them also have a higher probability to host the {\it same} or similar magnetic flux density.
Thus, it seems that the high occurrence in the boundary is merely due to the presence of long-lived magnetic elements piled up in relatively small regions.
If this were the case everywhere in the FoV, we should find an even steeper increase in $t_D$ as a function of the occurrence in the vertices of the supergranule where the strongest downflows are observed \citep{2014ApJ...788..137G} in correspondence with the strongest magnetic fluxes.
Instead, we found unequivocal signatures of a separated population of magnetic elements living in the vertices of the supergranule, and having a shorter $t_D$ (in green in Figure \ref{Fig.Regions} and in the middle right panel of Figure \ref{Fig.maps}) despite the maximum occurrence.
We exclude any effect due to low statistics since such a population is well-defined, localised, and represents a spatially correlated cluster, not randomly distributed throughout the FoV, which can be very unlikely produced by a noisy distribution of magnetic elements with constrained occurrence and decorrelation time.
These results suggest that different tightly packed magnetic elements in the vertices of the supegranule move in a very restricted region, thus allowing a high occurrence (near $100\%$) and $t_D\simeq40-50$ minutes.
This agrees with the low-diffusion, back-and-forth trajectories of magnetic elements found in the boundary of the supergranular cell, especially close to the vertices \citep{2013ApJ...770L..36G}, and is consistent with the horizontal velocity field found there and shown in the bottom panels of Figure \ref{Fig.maps}. 
A key role is probably played by intense downflows acting as attractors, constraining the dynamics of close magnetic elements to a random walk restricted to small regions, and causing magnetic elements to pile up there.


\section{Summary and conclusions}
Due to the lack of an exhaustive physical theory, turbulent convection in the quiet Sun can be approached by i) MHD simulations, ii) tracking observed magnetic elements, or iii) studying the observed spatial and temporal correlations of both plasma and magnetic features filling the FoV.
Here we followed the last approach, studying the occurrence and persistence of magnetic features in the quiet Sun.
The main results found can be itemised as follows:
\begin{itemize}
\item The occurrence analysis shows a pattern on supergranular spatial and temporal scales with occurrences ranging from $0\%$ in the central part of the supergranular cell to almost $100\%$ near its vertices.
\item The persistence analysis shows long-correlated patterns localised mainly in the supergranular boundary where decorrelation times up to $t_D\sim240$ minutes are found. 
\item The highest occurrences are found near the vertices of the supergranular cell, but in these regions $t_D$ lies in the middle of its observed range.
This suggests that the same locations are filled by different magnetic elements, probably due to the presence of intense downflows constraining their nearly random motion and acting as attractors. 
\item The increase in $t_D$ as a function of the occurrence steepens moving from the centre to the boundaries of the supegranule, indicating that magnetic elements in the boundaries move, on average, slower and are correlated longer with respect to those inside the supergranule.
The different slopes emerging from this behaviour may reflect the different magneto-convection regimes characterising the interaction between plasma and magnetic field in the quiet Sun.  
\end{itemize}

The occurrence and persistence analyses proposed in this work provide additional information on the dynamic constraints of the magnetic elements on single-pixel spatial scales, which make it difficult to distinguish between different magnetic elements with other methods and algorithms.
Further studies will be aimed to understand whether our findings are universal features characterising the solar quiet photosphere, or are strictly dependent on its physical conditions at the time when observations are carried out.
We also will extend our analysis to dopplergrams, investigating more extensively the onset of spatial and temporal correlations in magnetic features and their relationship with the local convection, with particular interest in the effect of the strongest downflows on the dynamics of magnetic elements. 

It is our opinion that studies on the scales of organisation observed on the Sun performed with statistical tools like the ones employed in this work can effectively gather information about the turbulent convection dynamics and that, in general, this approach will contribute significantly to shedding light on the debated aspects of convection in stellar atmospheres.

\begin{acknowledgements}
This research has received funding from the Italian MIUR-PRIN grant 2012P2HRCR on ``The active Sun and its effects on Space and Earth climate''.
Financial support from the Spanish Ministerio de Econom{\'i}a, Industria y Competitividad through projects ESP2014-56169-C6-1-R and  ESP2016-77548-C5-1-R, including a percentage from European FEDER funds, is gratefully acknowledged.
This paper is based on data acquired in the framework of the Hinode Operation Plan 151 entitled ``Flux replacement in the network and internetwork''.
Hinode is a Japanese mission developed and launched by ISAS/JAXA, collaborating with NAOJ as a domestic partner, and NASA and STFC (UK) as international partners. Scientific operation of the Hinode mission is conducted by the Hinode science team organised at ISAS/JAXA. This team mainly consists of scientists from institutes in the partner countries. Support for the post-launch operation is provided by JAXA and NAOJ(Japan), STFC (U.K.), NASA, ESA, and NSC (Norway). 
\end{acknowledgements}


\end{document}